\let\accentvec\vec
\documentclass[11pt]{llncs}

\let\vec\accentvec

\usepackage[utf8]{inputenc}
\usepackage[english]{babel}
\usepackage{amsmath}
\usepackage{amsfonts}
\usepackage{amssymb}
\usepackage{url}
\usepackage[pdftex]{graphicx}
\renewcommand{\tabcolsep}{2ex} 
\usepackage{subfigure}
\usepackage{booktabs}
\usepackage{cite}
\usepackage{fullpage}

\usepackage[colorlinks]{hyperref}
\usepackage[colorinlistoftodos]{todonotes}

\graphicspath{{img/}}

\DeclareMathOperator*{\argmax}{arg\,max} 

\newcommand{\CHordFunc}[1]{\ensuremath{\prec}}

\newcommand{\apath}[1]{\ensuremath{p_{#1}}} 				
\newcommand{\spath}[1]{\ensuremath{\mathcal{P}_{#1}}} 		
\newcommand{\brpath}[2]{\ensuremath{\left\langle{#1}, \ldots, {#2}\right\rangle}}
\newcommand{\length}[1]{\ensuremath{\mathcal{L}(#1)}}
\newcommand{\dist}[1]{\ensuremath{\mathcal{D}(#1)}}

\newcommand{\corridor}[2]{\ensuremath{\mathcal{C}^{#1}_{#2}}}	

\newcommand{\ie}{i.\@e.\@\xspace}

\title{Pruning Techniques for the Stochastic on-time Arrival Problem\texorpdfstring{\\}{ }--\texorpdfstring{\\}{ }An Experimental Study}
\author{Moritz Kobitzsch$^1$, Samitha Samaranayake$^2$, and Dennis Schieferdecker$^1$}
\institute{$^1$Karlsruhe Institute of Technology, $^2$University of California, Berkeley\\
\email{\{kobitzsch, schieferdecker\}@kit.edu, samitha@berkeley.edu}}

\begin{document}

\maketitle

\begin{abstract}
Computing shortest paths is one of the most researched topics in algorithm engineering.
Currently available algorithms compute shortest paths in mere fractions of a second on continental sized road networks.
In the presence of unreliability, however, current algorithms fail to achieve results as impressive as for the static setting.
In contrast to speed-up techniques for static route planning, current implementations for the \emph{stochastic on-time arrival} problem require the computationally expensive step of solving convolution products.
Running times can reach hours when considering large scale networks.
We present a novel approach to reduce this immense computational effort of stochastic routing based on existing techniques for alternative routes.
In an extensive experimental study, we show that the process of stochastic route planning can be speed-up immensely, without sacrificing much in terms of accuracy.
\end{abstract}

\section{Introduction}
\label{sec:intro}
Shortest path computation has come a long way since Edgar W. Dijkstra first formally introduced and solved the problem in 1959.
Modern speed-up techniques -- techniques that process road networks in a separate step under the assumption of a static network and augment them with additional information -- compute shortest paths on continental sized road networks in less than a millisecond.
Pure distance queries can be performed even faster.
However, these techniques do not consider the robustness of the found path but rather focus solely on the plain travel time.
This approach presents a severe shortcoming with regards to actual travel.
Any speed-up technique will ignore a path that takes a few seconds more, even though the shortest path might come with a high risk of traffic jams while the ``longer'' path is usually less frequented.
Two general approaches have been taken to counter these problems.
One method focuses on time-dependency: Instead of a plain travel time, streets are assigned a travel time function that depends on the actual arrival time.
As such, reoccurring phenomena as the morning or evening rush can usually be modeled quite well.
However, such approaches still only focus on an expected travel time and do not consider associated risks.
Another approach is to actually view the travel time as a random variable that behaves according to some probability distribution.
In this setting, one is interested in finding not only a reliable path but rather a full strategy that depends on a remaining time budget.
Finding such a strategy, described via a cumulative distribution function at every intersection, is expensive, though.
To compute the distributions, each step requires multiple convolutions.
In addition, even though it is known how to optimize the order of convolutions to minimize the computational effort~\cite{sbb-tcarrsn-12}, an optimal strategy might contain loops and with longer distances, the respective overhead grows immensely.
Running times up to hours are possible.

We present a new approach to account for the reliability in a more tractable manner.
As, similar to time dependent routing, reliability depends on measured data and deduced predictions, we propose to abandon guaranteed exactness.
In this work, we study different techniques that allow the computation of reliable shortest paths up to multiple orders of magnitude faster without a large impact on the actual quality of the strategies themselves; we do so by focusing on meaningful subsets of the road network based on recent research into alternative routes.

\section{Definitions and Terminology}
\label{sec:pre}
Every road network can be viewed as a directed and weighted graph. We formalize this notion in Definition~\ref{def:graph}.

\begin{center}
\begin{minipage}[c]{35em}
\begin{definition}[Graph, Restricted Graph]\label{def:graph}
A weighted \textbf{graph} $G = (V,A,c)$ is described as a set of vertices $V, |V| = n$, a set of arcs $A \subseteq V \times V, |A| = m$ and a cost function $c: A \mapsto \mathbb{N}_{>0}\text{.}$ For a given arc $(u,v)$, $c_{u,v}$ denotes the cost the arc (usually a form of travel time).
\end{definition}
\end{minipage}
\end{center}

All of the methods described within this work are targeted at the computation of shortest paths.
The required modelling of paths and their associated distances is summarized in Definition~\ref{def:paths}.

\begin{center}
\begin{minipage}[c]{35em}
\begin{definition}[Paths, Length, Distance]\label{def:paths}
Given a graph $G = (V,A,c)$:
We call a sequence $\apath{s,t} = \brpath{s=v_0}{v_k = t}$ with $v_i \in V, (v_i,v_{i+1}) \in A$ a \textbf{path} from $s$ to $t$.
Its \textbf{length} $\mathcal{L}(\apath{s,t})$ is given as the combined weights of the represented arcs: $\mathcal{L}(\apath{s,t}) = \sum_{i=0}^{k-1} c(v_i,v_{i+1})\text{.}$
If the length of a path $\apath{s,t}$ is minimal over all possible paths between $s$ and $t$ with respect to $c$, we call the path a \textbf{shortest path} and denote $\apath{s,t} = \spath{s,t}\text{.}$
The length of such a shortest path is called the \textbf{distance} between $s$ and $t$: $\dist{s,t} = \length{\spath{s,t}}$.
Furthermore, we define $\spath{s,v,t}$ as the \textbf{concatenated path} $\spath{s,v,t} = \spath{s,v} \cdot \spath{v,t}\text{.}$
\end{definition}
\end{minipage}
\end{center}

\section{The Stochastic on-time Arrival Problem (SOTA)}
\label{sec:sota}
Following, we give an overview over the current state of the art before describing our problem in detail.
We conclude the section by discussing a competing model to solve the problem.

\subsection{Related Work}
\label{sec:sota-related}
One approach to unreliable travel times is to take into account the inherent variability in the travel time distributions and compute robust routing strategies.
A commonly used method in this setting is to maximize the probability of reaching the destination within a prespecified time budget.
There are two different approaches that have been developed to solve this problem.
The first approach is to find the optimal \textit{a priori} path that maximizes the probability of on-time arrival and is known as the shortest path with on-time arrival reliability (SPOTAR) problem.
Nie and Wu~\cite{Nie2009} proposed a general solution to the problem based on first-order stochastic dominance of paths, but the solution has worst-case exponential running times.
They also present a pseudo-polynomial algorithm that gives an approximate solution to the problem and performs well in practice.
Nikolova et al.~\cite{Nikolova2006a} showed how to solve the problem in $n^{\Theta(\log n)}$ time when the arc travel time distributions are Gaussian.

The second approach is to find a routing policy that determines the optimal route by selecting the best next direction at each junction, and is known as stochastic on-time arrival (SOTA) problem.
Such a strategy can result in different paths based on the realized travel times at intermediate road segments and will provide a success probability that is greater than or equal to that of the SPOTAR path.
Fan et al.~\cite{fan} formulated the SOTA problem as a stochastic dynamic programming problem and solved it using a standard \textit{successive approximation} (SA) algorithm.
However, in a network that contains cycles as is the case with all road networks, there is no finite bound on the maximum number of iterations required for the algorithm to converge.
This is due to the fact that the optimal solution can contain loops, see \cite{sbb-tcarrsn-12} for an example.
As an alternative, Nie et al.~\cite{nie2006} proposed a discrete approximation algorithm for the SOTA problem which converges in a finite number of steps and runs in pseudo-polynomial time. Samaranayake et al.~\cite{sbb-tcarrsn-12} presented a number of optimization techniques to speed up the computation including a label-setting algorithm based on the existence of a uniform, strictly positive minimum arc travel time, advanced convolution methods centered on the Fast Fourier Transform and the idea of zero-delay convolution~\cite{sbb-spsoap}, and localization techniques for determining an optimal order of policy computation. The resulting algorithm is later referred to as \emph{classic} implementation.
Sabran et al.~\cite{sbb-stsoap-2014} showed how preprocessing techniques can be used to further reduce the computation time of the SOTA problem.
Unfortunately, the problem structure of the SOTA formulation limits the types of preprocessing methods that can be used for this problem, and prevents massive running times reductions as possible in the deterministic case.
Furthermore, this leads to large precomputation times due to the inability to decompose the respective tasks into smaller ones.

\subsection{Problem Definition}
\label{sec:sota-problem}

We consider the stochastic on-time arrival (SOTA) problem of finding the optimal routing strategy for reaching a given destination within a prespecified time budget.
In this problem, the costs of each arc $(u,v)\in{E}$ are random variables with a probability density function $c_{u,v}(\cdot)$ that represents the travel time on arc $(u,v)$.
The arc travel time distributions are assumed to be independent\footnote{See~\cite{sbb-tcarrsn-12} for an extension of the formulation that considers local correlations.}.
Given a time budget $T$, an \textit{optimal routing policy} is defined to be a routing policy that maximizes the probability of reaching the destination vertex $t$ within a total travel time of $T$.
A routing policy is an \textit{adaptive} set of instructions that determines the optimal path at each vertex (i.e., at an intersection in the road network) based on the cumulative travel time that has already been realized.
This is in contrast to the SPOTAR solution~\cite{Nie2009, Nikolova2006b} that only gives a fixed path from the source to the destination.

Given a vertex $u\in{V}$ and a time budget $\tau$, let $prob_{u,t}(\tau)$ denote the probability of reaching the destination $t$ from $u$ in less than time $\tau$ when following the optimal policy, and let $next_{u,t}(\tau)$ be the optimal next vertex to visit.
At each vertex $u$, the traveler should pick the arc $(u,v)$ that maximizes the probability of arriving on time at the destination.
If $v$ is the next vertex being visited after vertex $u$ and $\omega$ is the time spent on arc $(u,v)$, the traveler starting at $u$ with a time budget $\tau$ has a time budget of $\tau - \omega$ to travel from $v$ to the destination\footnote{In this formulation of the problem, the traveler is not allowed to wait at any of the intermediate vertices. See~\cite{sbb-tcarrsn-12} for the conditions under which travel time distributions from traffic information systems satisfy the first-in-first-out (FIFO) condition. This condition implies that the on-time arrival probability can not be improved by waiting at any vertex.}.

The optimal routing policy for the SOTA problem can be obtained by solving the following system of equations.
\begin{eqnarray}
\label{eq:sota}
prob_{u,t}(\tau)	& = & \max_{(u,v)\in{E}} \int_{0}^{\tau}c_{u,v}(\omega)\,prob_{v,t}(\tau-\omega)d\omega \\
				&   &	\hspace*{5em} \forall {u}\in{V}, \enspace u\neq{t}, \enspace 0\le{\tau}\le{T} \nonumber \\
prob_{t,t}(\tau) 	& = & 1	\hspace*{5em} 0\le{\tau}\le{T} \nonumber \\[1.5em]
next_{u,t}(\tau)	& = & \argmax\limits_{(u,v) \in E} \int_{0}^{\tau}c_{u,v}(\omega)prob_{v,t}(\tau-\omega)d\omega \\
				&   &	\hspace*{5em} \forall {u}\in{V}, \enspace u\neq{t}, \enspace 0\le{\tau}\le{T} \nonumber
\end{eqnarray}
One approach to solving this problem would be to use a \textit{successive approximations} (SA) algorithm as in~\cite{fan}, which solves the system of equations~(\ref{eq:sota}) repeatedly until convergence and gives an optimal routing policy. If the arc travel time distributions are lower bounded by a strictly positive constant and uniformly bounded on the network, as is the case with road networks, the system can be solved more efficiently using a label setting algorithm~\cite{sbb-tcarrsn-12}.

\subsection{A Different Approach}
\label{sec:time-dependent-related}
Another approach to model variations in travel time along a road segment is to consider time-dependent routing.
In the time-dependent scenario, arcs are assigned a travel time function that depends on the actual time of arrival.
Dreyfus \cite{d-aspa-69} gives an early result to solve earliest arrival queries using a modified version of Dijkstra's algorithm.
Orda and Rom \cite{or-smantd-90} discuss different variations of potential \emph{travel time functions} (TTF), showing that the FIFO property can be achieved for functions violating it using an auxiliary TTF.
Fast algorithms exist to solve earliest arrival queries \cite{d-tdsharc-11} and profile queries \cite{bgsv-mtdttch-13} on such networks; a profile query gives a travel time profile for a full period of time.
Nevertheless, the SOTA problem remains an interesting problem to solve as the types of delays that can be modeled using travel time functions for time-dependent routing differ from the types of uncertainty modeled in travel-time distributions.
Typical approaches for time-dependent routing are regular and predictable delays. Potential examples include roads that have different speed-limits at night or the morning/evening rush.
In the stochastic setting, one is more interested in irregular delays that are not predictable.

\section{Pruning Techniques for the SOTA Problem}
\label{sec:pruning}
In a basic SOTA query, most of the road network $G$ only contributes a tiny fraction to the PDF at each vertex.
Therefore, it is reasonable to limit the search space of each query to a subgraph $G'$ that allows us to compute near optimal results with respect to the PDFs at a fraction of the normally required convolutions.
We conjecture that paths that are contributing a significant amount to the PDFs are likely to be part of a reasonable alternative route between source $s$ and destination $t$.
Our results in the experimental section show this conjecture to be sound.
Following, we give a short recapitulation on previous work on alternative routes, before describing required adjustments to these original methods.

\subsection{Existing Techniques}
The following techniques are designed for static route planning. In the stochastic setting, we evaluate distributions for their mean to apply these techniques.

\paragraph*{Via-Node Alternative Routes.}
The via-node approach to alternatives routes was first introduced in \cite{c-cr-05} and later refined by \cite{adgw-arrn-13,ls-csarr-14,k-hidar-13}.
The main idea behind this approach is to model alternative routes $\spath{s,v,t}$ as a concatenation of two shortest paths $\spath{s,v}$, $\spath{v,t}$, with $v$ called a \emph{via node}.
A candidate set of via nodes $V_C$ can be computed by intersecting the (pruned) search spaces from $s$ in forward direction and from $t$ in backward direction.
The actual via node is selected by some feasibility criteria elaborated in the original work.

\paragraph*{Penalty Method.}
The penalty method \cite{bdgs-argrn-11,krs-eepmag-13,pz-iarp-13} considers a different approach to alternative routes that allows us to find routes of a different structure than the via-node approach.
It constructs a set of alternative routes between $s$ and $t$ by iteratively computing shortest paths between these two vertices and penalizing the arcs on this path as well as the adjoint arcs.
The iteration stops once the found paths get too long or after a set number of rounds.
A path computed in this fashion is only retained as an alternative route if it adheres to some quality criteria discussed in the original publications.
The subgraph induced by the set of found alternative routes and the shortest path is called an \emph{alternative graph}.

\paragraph*{Route Corridors.}
A related problem is considered in \cite{dklw-rmrplc-12}.
Here, the goal is to compute a corridor of paths from $s$ to $t$ that contain all arcs that are possibly visited when driving from $s$ to $t$ and performing up to $k$ successive wrong turns.
We call this subgraph of $G$ a \emph{$k$-turn corridor} $\corridor{k}{s,t}$.
It can be constructed recursively as $\corridor{k}{s,t}$ is the union of all shortest paths $\spath{v,t}$ with $v \in \corridor{k-1}{s,t}$ and $\corridor{0}{s,t} = \spath{s,t}$.

\subsection{Adjustments for Our Setting}
As the experimental results show most of these techniques perform remarkably well right out of the box.
The penalty method and route corridors directly yield a subgraph $G'$ that we can use to prune the SOTA search. Although, in our setting we keep all paths the penalty method finds and do not evaluate for quality.
For the via-node approach, however, we have to come up with a new solution for constructing $G'$, as the basic via-node approach only yields single alternative routes.
It can be iterated to generate additional routes that are an alternative to the shortest path and to all previous alternative routes, though, and we could use the union of these routes as an alternative graph similar to the penalty method.
However, only few routes are found on average leading to a very sparse graph $G'$.
We therefore opt for another option.
We take the union of all paths $\spath{s,v,t}$ with $v \in V_C$ that adhere to the feasibility criteria as alternative graph.
This corresponds to an approach previously discussed in \cite{k-hidar-13}.

\section{Evaluation Framework}
In this work, we present an experimental study on pruning techniques.
The relevance of these techniques strongly depends on the actual distributions found in the network.
We present values based on a variety of different inputs, ranging from measured data incorporated into traffic models to complete random data.
The focus of our experiments is San Francisco, for which measured data is available to us. In addition, we present measurements on randomly generated gamma distributions \cite{KapariasBellBelzner08} for San Francisco as well as the district of Karlsruhe.

\subsection{Measured Data}
Measured data available to us is due to the Mobile Millennium Project (\url{http://traffic.berkeley.edu/}).
The project collects GPS based probe data from a number of sources and processes this data using filtering and estimation techniques~\cite{hunter12wafr} to obtain travel-time distributions at the granularity of road segments. The resulting Mobile Millennium input set describes possible travel times in San Francisco as normal mixtures.
In a smoothed setting we scale the parameters of extreme distributions (see Figure~\ref{fig:variance-network}).

\subsection{Generated Data}
We consider three variations of path based randomness as well as two path independent settings.
Except for a completely random setting, all of the techniques operate in multiple rounds.
The characteristic of the resulting graphs is described in Table~\ref{tab:input-char}.

For the path based settings, we compute a range of different paths each round, following one of the three paradigms specified below.
During each round, we keep a counter for every arc that is incremented whenever an arc is part of a shortest path. 
Based on these counters, we increase (or decrease) the respective parameters of the gamma distributions (shape and scale) of the associated arc.
For a decrease we have chosen a multiplicative factor of 0.9, for increase we have chosen additive penalties of 0.02/0.08 for shape and 0.1/0.5 for scale.
The resulting distributions somewhat resemble distributions found in the measured data.
The individual path based settings operate as follows:

\paragraph*{Random Paths:}
For the random paths setting, counters are based on source and target pairs chosen uniformly at random.
After a shortest path computation using \emph{free flow times} (best possible travel time), the counters of the arcs on the respective path are increased.

\paragraph*{Random Shuffle:}
Instead of choosing source and target pairs uniformly at random, we pair each vertex of the graph with another random vertex of the graph. This results in every vertex acting as source and target vertex once in every round. As such, the number of paths considered is given as the number of vertices in the graph.

\paragraph*{Hotspots:}
The hotspot setting is similar to the random path setting but creates more localized penalties.
In every round, a few random hotspots are chosen. The paths to compute are distributed equally over these hotspots.
For every such path, we choose a source vertex uniformly and compute the path to the desired hotspot.\\

The non path based settings operate as follows:

\paragraph*{Random Arcs:}
A completely random setting for which we simply increase the count of randomly chosen arcs. In this setting, we interpret each arc as a one-hop path and operate as described above.

\paragraph*{Random Arc Distributions:}
As a final test of robustness, we also generated a completely unrealistic setting: random distributions.
In this setting we do not operate in rounds but rather assign a random gamma distribution with shape $\in \left[ 0.01, 10 \right]$ and scale $\in \left[0.01, 10\right]$ to each arc directly.

\begin{table}
	\centering
	\caption{Characteristics of our main inputs. Setting gives appropriate parameters for \#(R)rounds, \#(P)aths, and \#(H)otspots. Order describes the relation between the classic/optimal ordering; optimal ordering is an a posteriori ordering that first performs an instance of the classic algorithm, extracts the actually contributing vertices, and prunes the search space to contain only said vertices. Type names the kind of distributions used (NM: Normal Mixtures or GD: Gamma Distribution). Nodes [\%] give the fragment of the input graph that makes up the optimal prune space (nodes that make up the perfect strategy after running the classic approach). Time budget by arrival probability describes the factor in relation to free-flow traffic that is necessary to guarantee on-time arrival with X\% probability. The graphs are illustrated in Figure~\ref{fig:variance-network}.}
	\label{tab:input-char}
	\begin{tabular}{clcccccccc}
\toprule
   &   &         &     &       &       & \multicolumn{4}{c}{Budget by Arr. Prob.}\\\cmidrule{7-10}
\# & Source & Setting & Type & Order & Nodes & 25 & 50 & 75 & 100 \\
\midrule
1 & Random Shuffle & R15 & GD & 2.22 & 16.78 & 1.08 & 1.11 & 1.14 & 1.45
\\
2 & Random Arcs & R40, P50k & GD & 3.34 & 14.86 & 1.11 & 1.14 & 1.18 & 1.73
\\
3 & Hotspot & R15, P5k, H50 & GD & 1.89 & 26.24 & 1.14 & 1.19 & 1.24 & 1.84
\\
4 & Random Paths & R100, P2.5k & GD & 3.97 & 19.27 & 1.29 & 1.37 & 1.47 & 2.19
\\
5 & Random Paths & R25 P10k & GD & 1.51 & 46.05 & 1.37 & 1.47 & 1.57 & 2.79
\\
6 & Mobile Millen. & smoothed & NM & 1.74 & 34.64 & 1.24 & 1.31 & 1.38 & 2.03
\\
7 & Mobile Millen. & - & NM & 1.33 & 56.24 & 1.45 & 1.58 & 1.73 & 3.86
\\
\bottomrule
	\end{tabular}
\end{table}

\section{Experiments}
\label{sec:exp}
Even though we present an experimental approach to speed-up stochastic-routing, we do not focus on running times.
Instead, we mostly rate the success of our approach via the reduction in the number of convolutions necessary to compute an optimal strategy.
Our decision to do so is based on the large influence of the multitude of design decisions necessary in implementing an algorithm to solve the SOTA problem.
Especially the method chosen to perform the convolution, trade-offs between memory overhead and pre-evaluated arc distributions, and memory allocation itself during the execution of the algorithm have a large influence on the running time; some of which might not be viable due to the associated overheads.
The main workload, however, remains in the cost of the convolutions.
By purely focusing on the plain number of convolutions performed, we give a less biased view on the benefits of our approach.
The running time of the used techniques is negligible, as all of them have been proven highly efficient and require only a few milliseconds on continental sized networks.

\paragraph*{Methodology:}
The data we present is measured on the small network of San Francisco. The graphs consists of 2.3k nodes.
Depending on the budget, however, running times can still reach minutes for the classic techniques as the optimal node order might still reach into the millions of steps.
Unless otherwise noted, we present data averaged over a thousand queries, source and target chosen uniformly at random.

\subsection{Absolute Errors}
\begin{figure}[tb]
	\subfigure[Mobile Millenium (Graph 7)]{\includegraphics[width=0.47\textwidth,page=3]{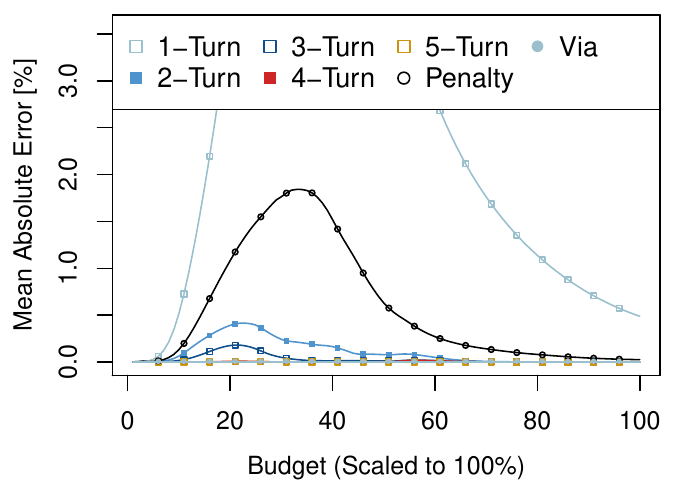}}\hfill
	\subfigure[Random Arcs (Graph 2)]{\label{fig:errors-cc}\includegraphics[width=0.47\textwidth,page=3]{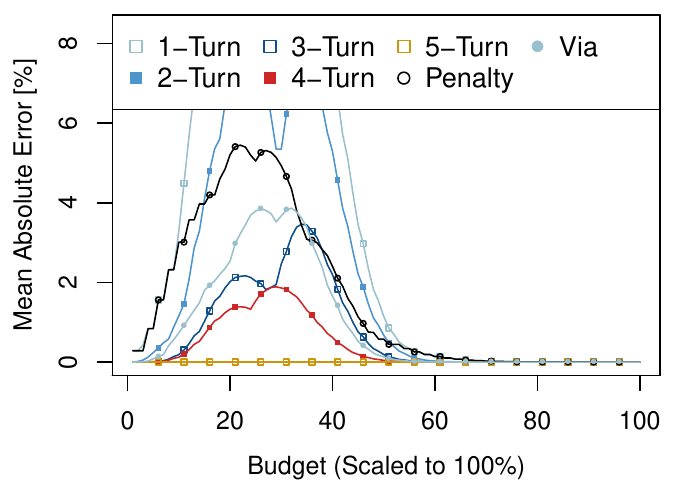}}\\
	\subfigure[Hotspot (Graph 3)]{\includegraphics[width=0.47\textwidth,page=3]{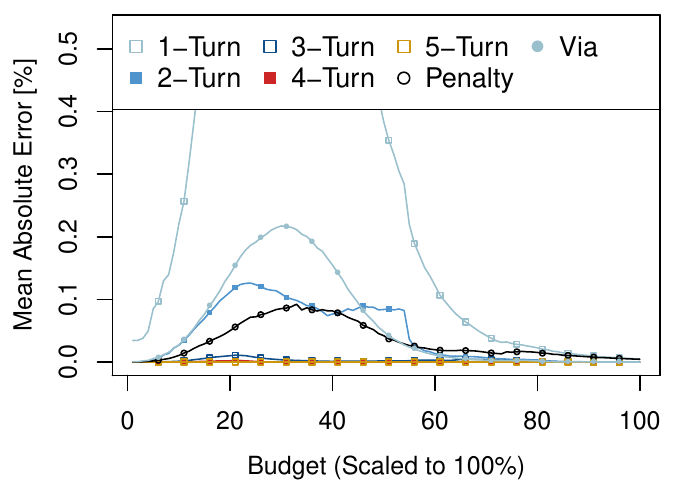}}\hfill
	\subfigure[Random Paths (Graph 4)]{\includegraphics[width=0.47\textwidth,page=3]{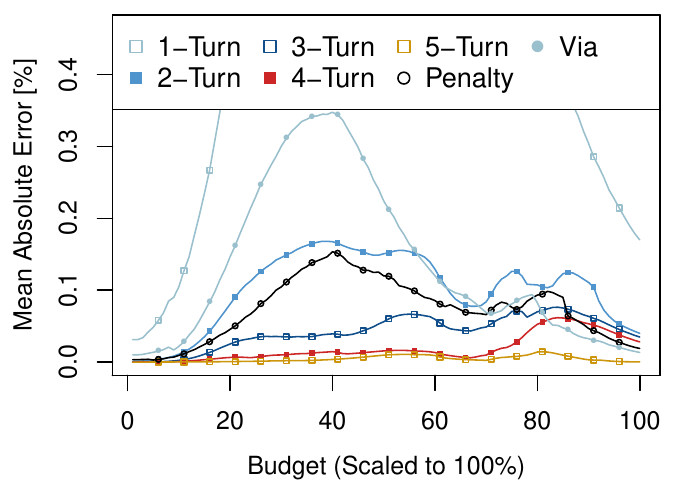}}
	\caption{Mean error in relation to given time budget, scaled to 100~\%. Pruning techniques evaluate for the expected mean travel time of an arc. Plots are scaled to a 100~\% between first relevant budget (arrival probability $>$ 0.001) and first 100~\% arrival guarantee.}
	\label{fig:pdf-errors-examples}
\end{figure}

In Figure~\ref{fig:pdf-errors-examples}, we present the absolute delta (error) of the pruned PDFs in relation to the unpruned PDF and to the relevant time budget over a selection of different graphs. While the relative order of the different techniques varies a bit, the experienced behavior more or less remains the same.
A $1$-turn corridor usually presents a viable method for most of the budget already, especially in the high budget regions, however, it might not suffice in terms of approximation quality as errors get relatively large for lower budget ranges. Higher degree corridors come closer to the optimal solution, close to a perfect approximation.

The quality of the approximation seems relatively stable.
It offers some higher maximal possible errors on a few graphs with difficult to handle extreme variances, though.
In the presence of extreme variance, allowing for high probabilities for travel times of more than 10 times the free flow time, the exclusion of a single arc can sometimes introduce errors as high as 10-15~\% in the PDF of the source vertex for various techniques. Even on these graphs, these kind of paths occur rarely and can be deemed unrealistic. A scenario that has to consider more than three times the travel time during free flowing traffic for a routing strategy seems unreasonable to the authors.

\paragraph*{Range Influence:}
In road networks, the typical topology of a shortest path changes with longer distance. Paths of sufficient lengths are usually restricted to a rather sparse network after leaving the respective region of the starting point or when far enough away from the target \cite{bfmss-itcsp-07}. To make sure our algorithm performs well for both long range and short range queries, we performed tests on different Dijkstra ranks on a larger network, consisting of the district of Karlsruhe; the Dijkstra rank of a node is the iteration in which Dijkstra's algorithm would settle the node. The graph consists of 129k nodes and 320k edges and we used the random path setting for generating distributions as it seems closest to the Mobile Millennium framework (see Table~\ref{tab:input-char}).
While we did experience some minor variation, the general result remained the same over all Dijkstra ranks.

\subsection{Technique benefits}
\begin{figure}[tb]
	\includegraphics[width=1.0\textwidth]{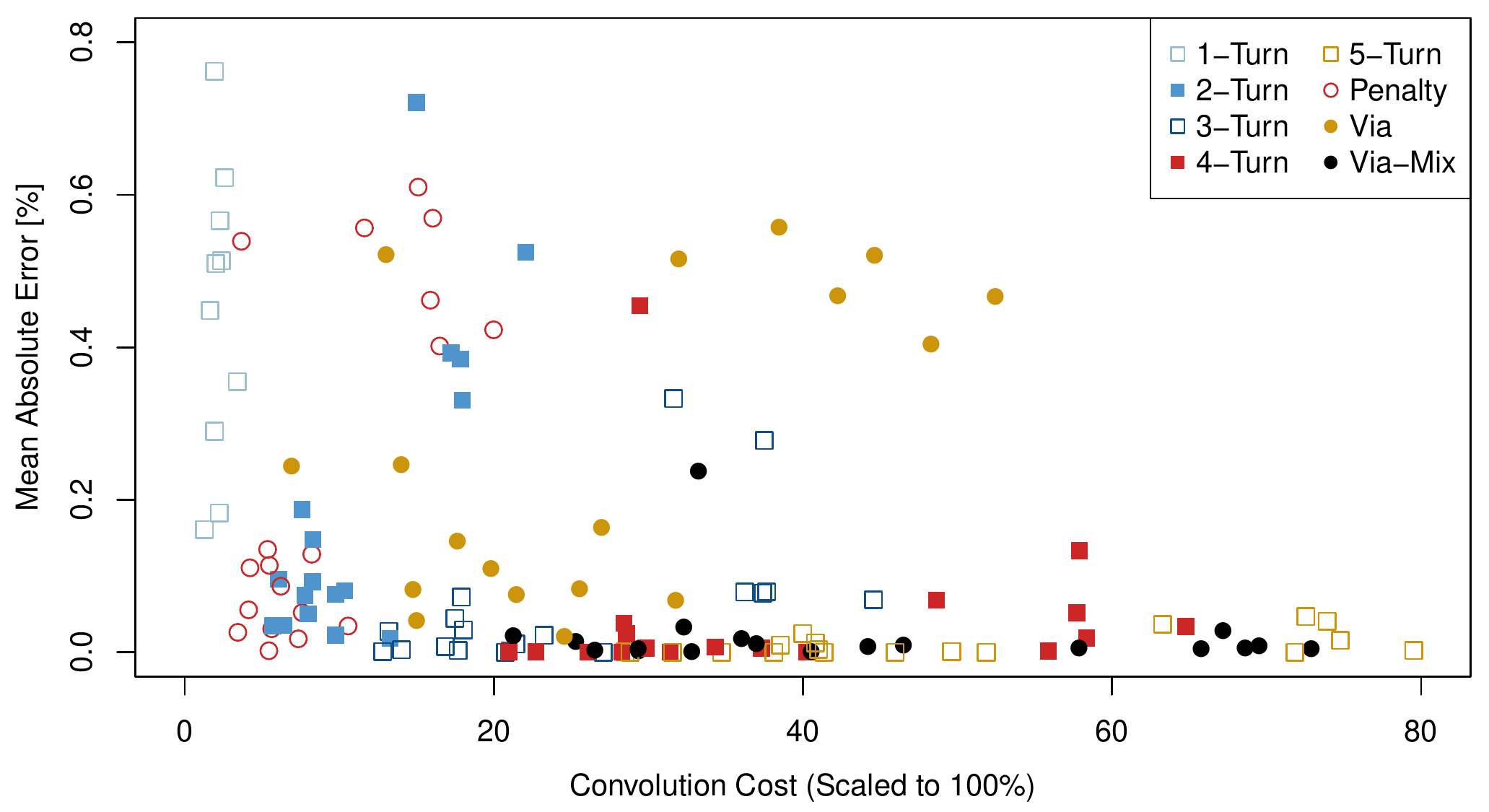}
	\caption{Number of convlutions in optimal SOTA-ordering and corresponding error in relation to the classic ordering and the maximal encountered error. Every dot represents a different input graph (e.g. mobile millenium, hotspots, random paths, random arcs) with varying parameters. Scaled to 100~\% in relation to the classic ordering.}
	\label{fig:scatter-overhead}
\end{figure}

The different pruning techniques not only vary in terms of their approximation quality but also in terms of their pruning quality.
We compare the differences in Figure~\ref{fig:scatter-overhead}. Every dot represents a different input graph, including additional graphs (in relation to Table~\ref{tab:input-char}) with different randomization parameters.

For the most part, both the Penalty method as well as the 2-turn corridor offer a reasonable trade-off between average experienced approximation rate and reduction in the number of convolutions. The 5-turn corridor does prune far less nodes, but also offers near perfect approximation.
The combination of alternative routes while evaluating arcs to their minimal, average, and maximal possible travel time (\emph{via-mix}) performs comparable to a 5-turn corridor.
Even though the reduction in the size of the node order varies between $5~\%$ and up to $78~\%$ of the node order required in the classic approach, this reduction has a far larger influence as one might expect at first glance.
Considering that the average node has a total of two to four reachable neighbors, the reduction in the ordering process translate into an even larger speed-up as every update may require four convolutions.
In addition, the pruned implementations usually prune some of these convolutions due to neighboring nodes not being considered at all, only performing two instead of four convolutions.
As such, we expect (and experienced in our implementation) possible speed-ups of more than one or two orders of magnitude.
The benefit, however, lessens when influencing factors like evaluation of arc distributions and memory allocation as well as a potential faster convolution implementation are considered.

\section{Conclusion and Future Work}
\label{sec:con}
This experimental study proves a high potential for utilizing alternative route techniques for difficult to solve problems on road networks, such as the stochastic on-time arrival problem. We show that concentrating on a meaningful subgraph of the network helps to reduce the calculation overheads immensely.
Using techniques as we present here, reductions by multiple orders of magnitude are possible, making it possible to handle relatively large networks of more than a hundred thousand vertices.
We expect the impact of our proposed pruning methods to be even more important on larger networks that require large budgets and contain large amounts of vertices.
Even though we verified our expectations in some sense on artificial inputs, it would be most important to gain access to actual measured distributions on larger instances to confirm our approach.
With access to more realistic data, it would be interesting to fit the pruning techniques to the cause to minimize maximal errors and to reduce the computational overhead at the same time. Potential improvements could focus on different methods based on the distance from source and target or budget dependent pruning as minimal and maximal necessary time budgets can be computed in a trivial fashion.
We expect turn corridors that allow for more turns at the start/end of the route to offer great results, both with respect to speed-up and to experienced errors.

\subsubsection*{Acknowledgements.}
We would like to thank Mehrdad Niknami for providing us with the graph of the Mobile Millennium framework.

\bibliographystyle{splncs03}
\bibliography{sota-pruning-tr}

\begin{thebibliography}{10}
\providecommand{\url}[1]{\texttt{#1}}
\providecommand{\urlprefix}{URL }

\bibitem{adgw-arrn-13}
Abraham, I., Delling, D., Goldberg, A.V., Werneck, R.F.F.: Alternative routes
  in road networks. In: Festa, P. (ed.) SEA. pp. 23--34. LNCS, Springer,
  Heidelberg (2010)

\bibitem{bdgs-argrn-11}
Bader, R., Dees, J., Geisberger, R., Sanders, P.: Alternative route graphs in
  road networks. In: Marchetti-Spaccamela, A., Segal, M. (eds.) TAPAS. pp.
  21--32. Springer, Heidelberg (2011)

\bibitem{bfmss-itcsp-07}
Bast, H., Funke, S., Matijevic, D., Sanders, P., Schultes, D.: In transit to
  constant time shortest-path queries in road networks. In: ALENEX. SIAM,
  Philadelphia (2007)

\bibitem{bgsv-mtdttch-13}
Batz, G.V., Geisberger, R., Sanders, P., Vetter, C.: Minimum time-dependent
  travel times with contraction hierarchies. ACM Journal of Experimental
  Algorithmics  18 (2013)

\bibitem{c-cr-05}
{Cambridge Vehicle Information Tech. Ltd}: Choice routing (2005),
  \url{www.camvit.com}

\bibitem{d-tdsharc-11}
Delling, D.: Time-dependent sharc-routing. Algrthmica  60(1),  60--94 (2011)

\bibitem{dklw-rmrplc-12}
Delling, D., Kobitzsch, M., Luxen, D., Werneck, R.F.F.: Robust mobile route
  planning with limited connectivity. In: Bader, D.A., Mutzel, P. (eds.)
  ALENEX. pp. 150--159. SIAM, Philadelphia (2012)

\bibitem{d-aspa-69}
Dreyfus, S.E.: An appraisal of some shortest-path algorithms. Oper. Res.
  17(6),  395--412 (1969)

\bibitem{fan}
Fan, Y., Nie, Y.: Optimal routing for maximizing the travel time reliability.
  NSpEc  6(3-4),  333--344 (2006)

\bibitem{hunter12wafr}
Hunter, T., Abbeel, P., Bayen, A.M.: The path inference filter: Model-based
  low-latency map matching of probe vehicle data. In: Siciliano, B., Khatib, O.
  (eds.) Algorithmic Foundations of Robotics X. pp. 591--607. STAR, Springer,
  Heidelberg (2012)

\bibitem{KapariasBellBelzner08}
Kaparias, I., Bell, M.G.H., Belzner, H.: A new measure of travel time
  reliability for in-vehicle navigation systems. J. Intell. Transport. S.
  12(4),  202--211 (2008)

\bibitem{k-hidar-13}
Kobitzsch, M.: An alternative approach to alternative routes: Hidar. In:
  Bodlaender, H.L., Italiano, G.F. (eds.) ESA. pp. 613--624. Lecture Notes in
  Computer Science, Springer, Heidelberg (2013)

\bibitem{krs-eepmag-13}
Kobitzsch, M., Radermacher, M., Schieferdecker, D.: Evolution and evaluation of
  the penalty method for alternative graphs. In: ATMOS. OASICS, vol.~33, pp.
  94--107. Schloss Dagstuhl - Leibniz-Zentrum fuer Informatik, Dagstuhl (2013)

\bibitem{ls-csarr-14}
Luxen, D., Schieferdecker, D.: Candidate sets for alternative routes in road
  networks. In: Klasing, R. (ed.) SEA. pp. 260--270. LNCS, Springer, Heidelberg
  (2012)

\bibitem{nie2006}
Nie, Y., Fan, Y.: Arriving-on-time problem. Transport. Res. Rec.  1964,
  193--200 (2006)

\bibitem{Nie2009}
Nie, Y., Wu, X.: Shortest path problem considering on-time arrival probability.
  Transport. Res. B-Meth.  43(6),  597 -- 613 (2009)

\bibitem{Nikolova2006b}
Nikolova, E., Brand, M., Karger, D.R.: Optimal route planning under
  uncertainty. In: Long, D., Smith, S.F., Borrajo, D., McCluskey, L. (eds.)
  ICAPS. pp. 131--141. AAAI, Palo Alto (2006)

\bibitem{Nikolova2006a}
Nikolova, E., Kelner, J.A., Brand, M., Mitzenmacher, M.: Stochastic shortest
  paths via quasi-convex maximization. In: Azar, Y., Erlebach, T. (eds.) ESA,
  pp. 552--563. LNCS, Springer, Heidelberg (2006)

\bibitem{or-smantd-90}
Orda, Rom: Shortest-path and minimium-delay algorithms in networks with
  time-dependent edge-length. J. {ACM}  37,  607--625 (1990)

\bibitem{pz-iarp-13}
Paraskevopoulos, A., Zaroliagis, C.D.: Improved alternative route planning. In:
  Frigioni, D., Stiller, S. (eds.) ATMOS. OASICS, vol.~33, pp. 108--122.
  Schloss Dagstuhl - Leibniz-Zentrum fuer Informatik, Dagstuhl (2013)

\bibitem{sbb-stsoap-2014}
Sabran, G., Samaranayake, S., Bayen, A.M.: Precomputation techniques for the
  stochastic on-time arrival problem. In: McGeoch, C.C., Meyer, U. (eds.)
  ALENEX. pp. 138--146. SIAM, Philadelphia (2014)

\bibitem{sbb-tcarrsn-12}
Samaranayake, S., Blandin, S., Bayen, A.: A tractable class of algorithms for
  reliable routing in stochastic networks. Transport. Res. C-Emer.  20(1),
  199--217 (2012)

\bibitem{sbb-spsoap}
Samaranayake, S., Blandin, S., Bayen, A.M.: Speedup techniques for the
  stochastic on-time arrival problem. In: Delling, D., Liberti, L. (eds.)
  ATMOS. pp. 83--96. OASICS, Schloss Dagstuhl - Leibniz-Zentrum fuer
  Informatik, Dagstuhl (2012)

\end{thebibliography}

\clearpage
\begin{appendix}
\section{Graph Illustration}
\begin{figure}[h]
	\subfigure[Graph 1]{\includegraphics[width=0.47\textwidth]{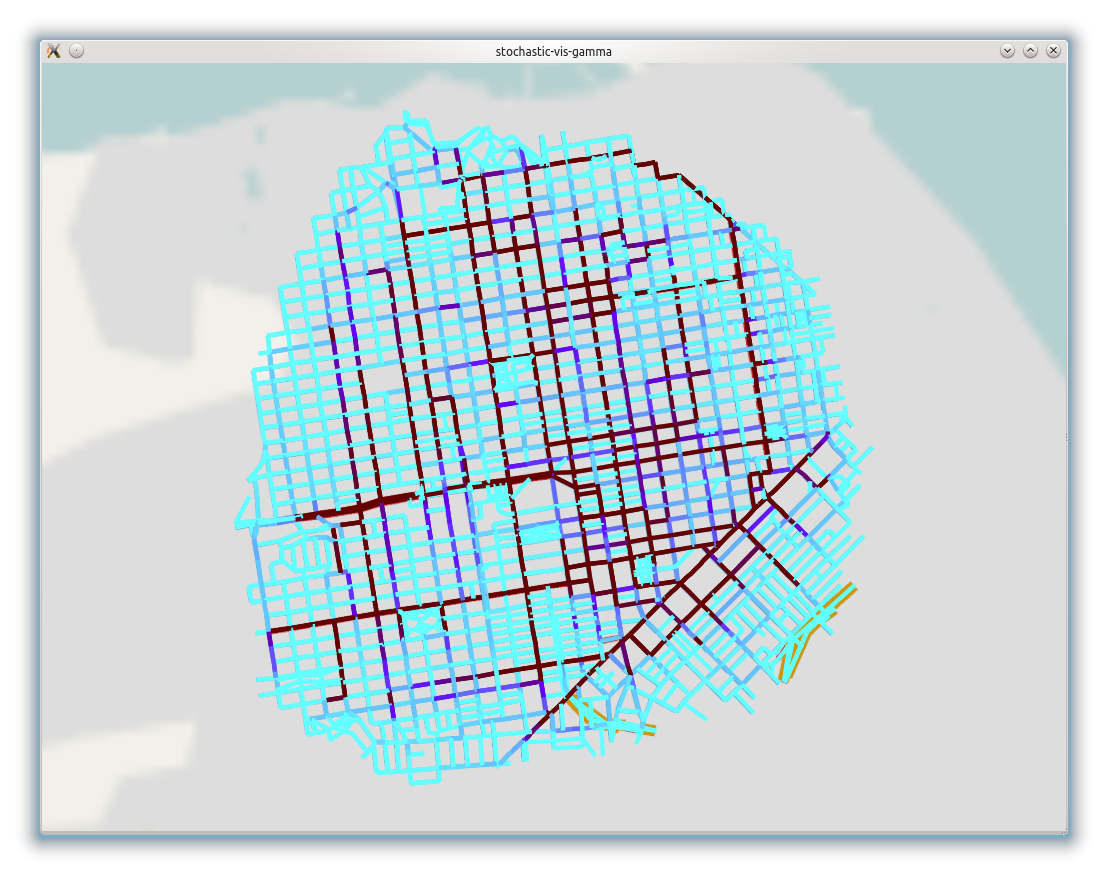}}\hfill
	\subfigure[Graph 2]{\includegraphics[width=0.47\textwidth]{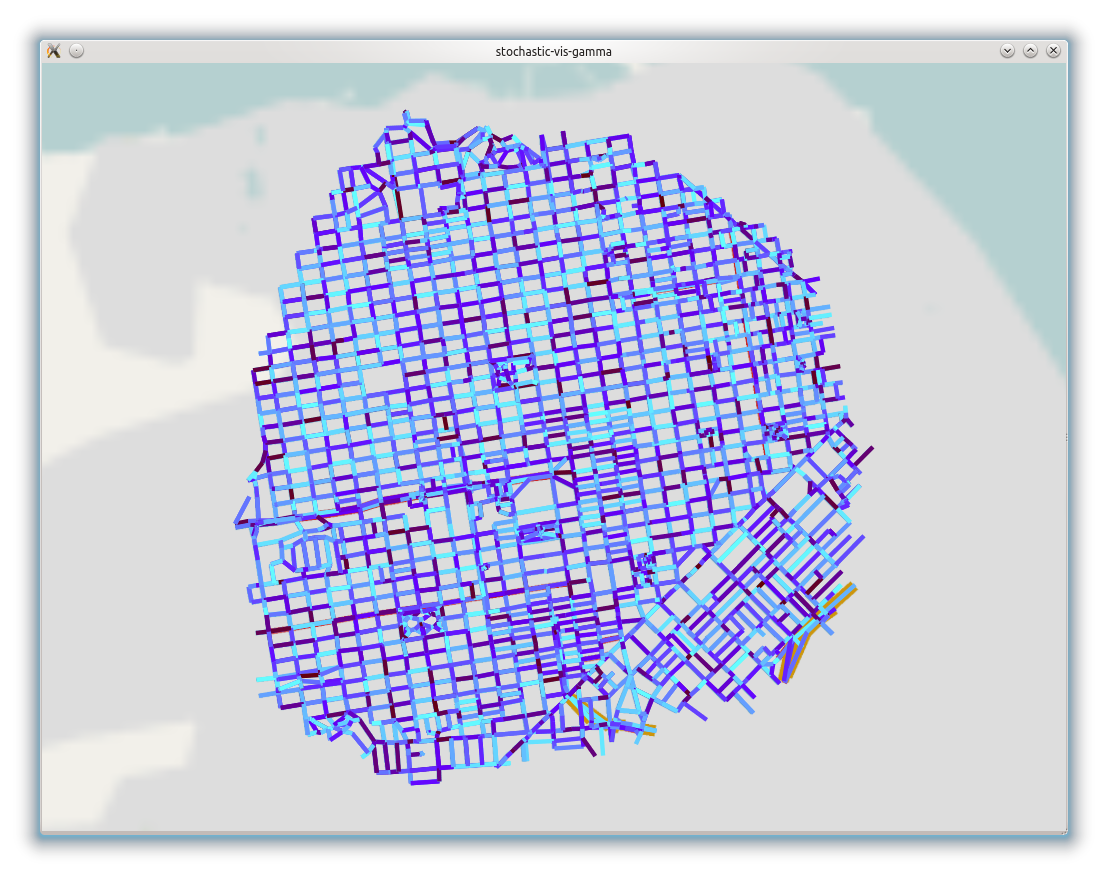}}\\
	\subfigure[Graph 3]{\includegraphics[width=0.47\textwidth]{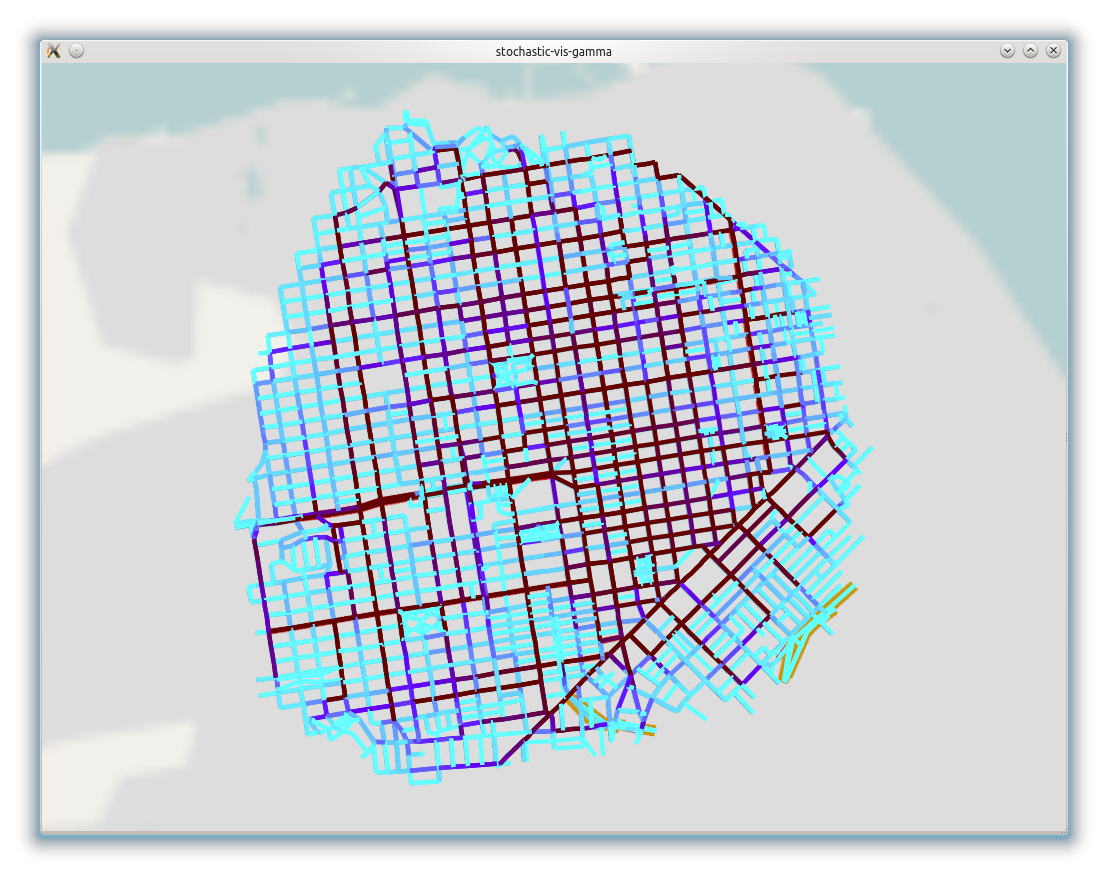}}\hfill
	\subfigure[Graph 4]{\includegraphics[width=0.47\textwidth]{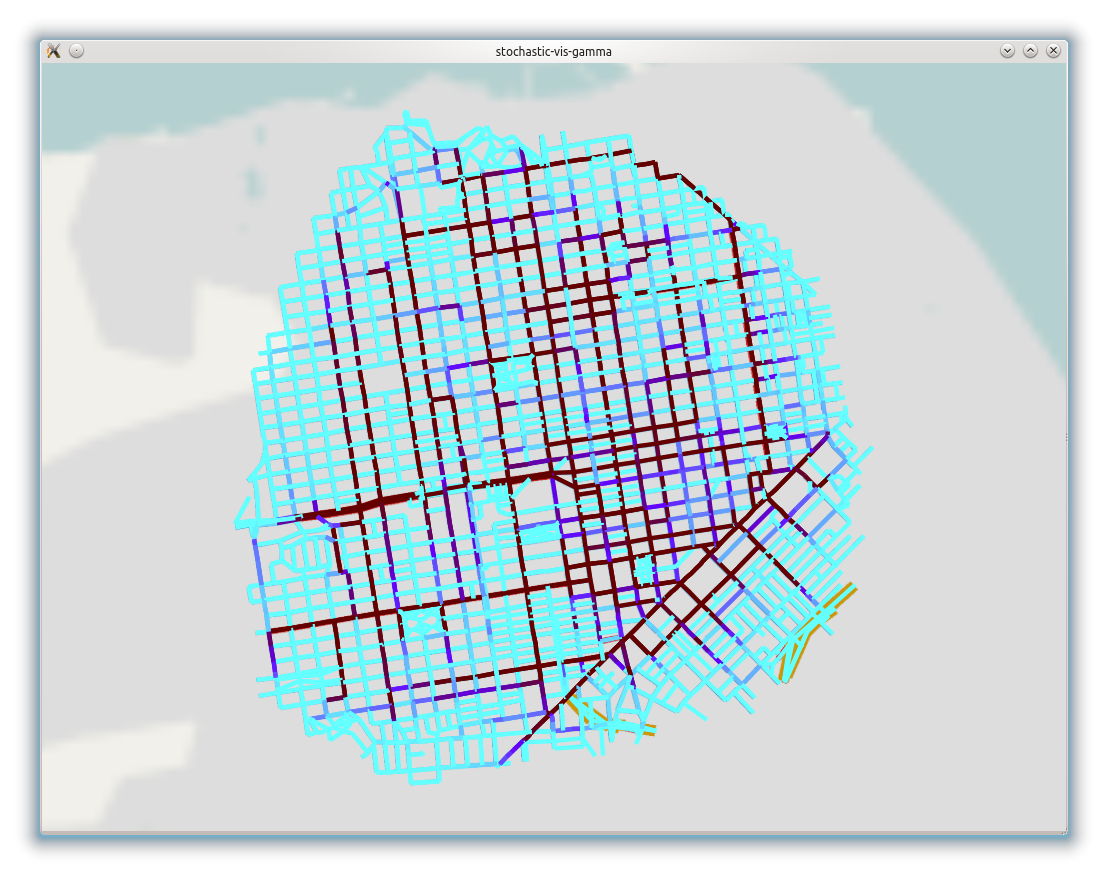}}
	\caption{Illustration of Arcs with high variance on the network. The darker the color of an arc, the more variance is associated with it. Given in relative scale from 0~\% (light blue) to 100~\% (dark red). Graph names are from Table~\ref{tab:input-char}.}
	\label{fig:variance-network}
\end{figure}

To better illustrate the different input, we present a graphical representation of the graphs in Figure~\ref{fig:variance-network}. The figure shows the relative variance assigned to the different streets. The minimal variance is shown in light blue, the maximal variance in dark red.
The difference between Graph 6 and Graph 7 (Table~\ref{tab:input-char}) shows the extreme variances present in the measured inputs on some arcs.
In relation to these extreme arcs, nearly no variance seems to be present on the network.
Smoothing these outliers reveals the variances on the other arcs.

\begin{figure}[h]
	\subfigure[Graph 5]{\includegraphics[width=0.47\textwidth]{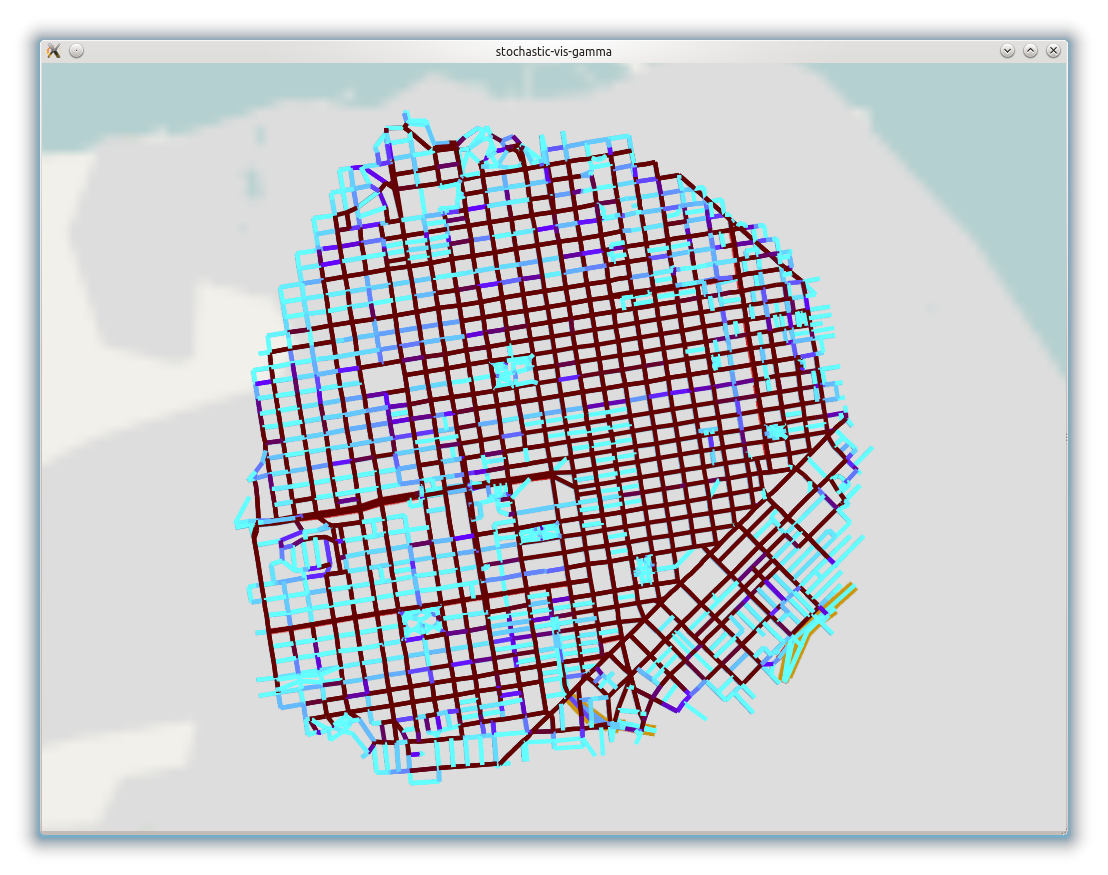}}\hfill
	\subfigure[Graph 6]{\includegraphics[width=0.47\textwidth]{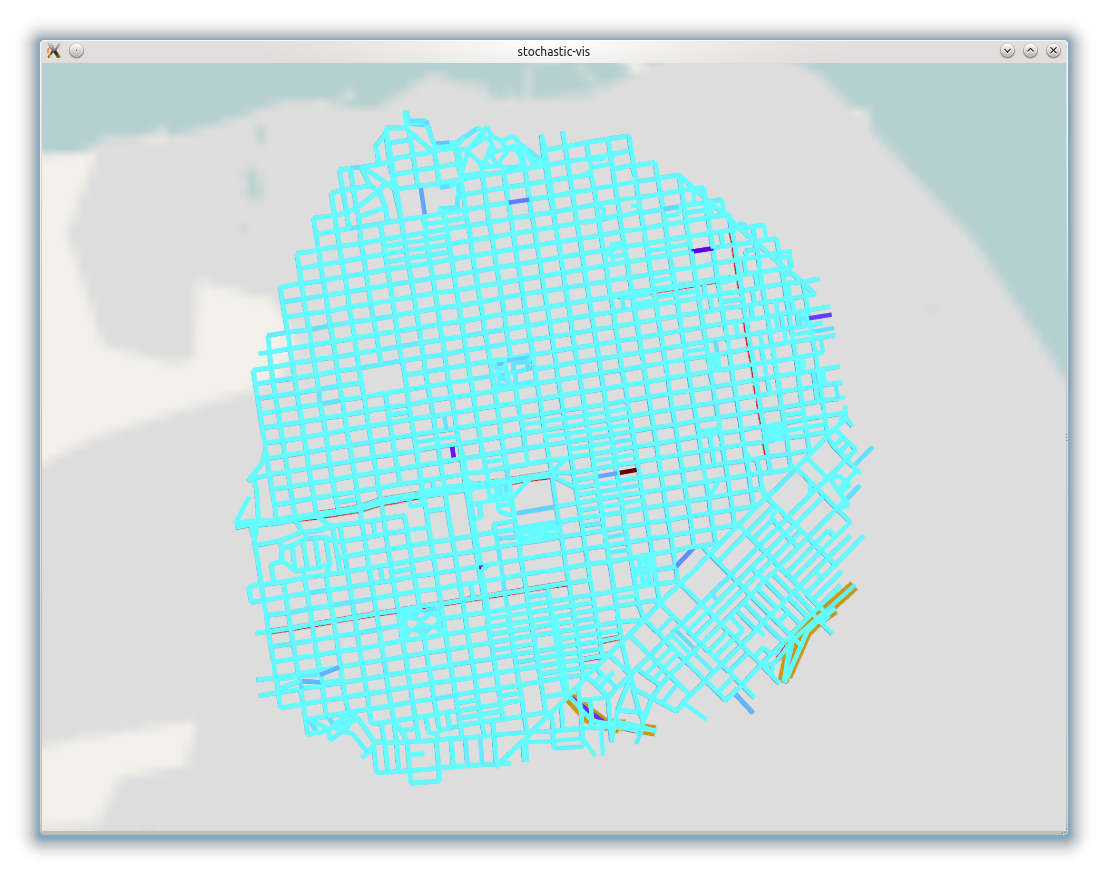}}\\
	\subfigure[Graph 7]{\includegraphics[width=0.47\textwidth]{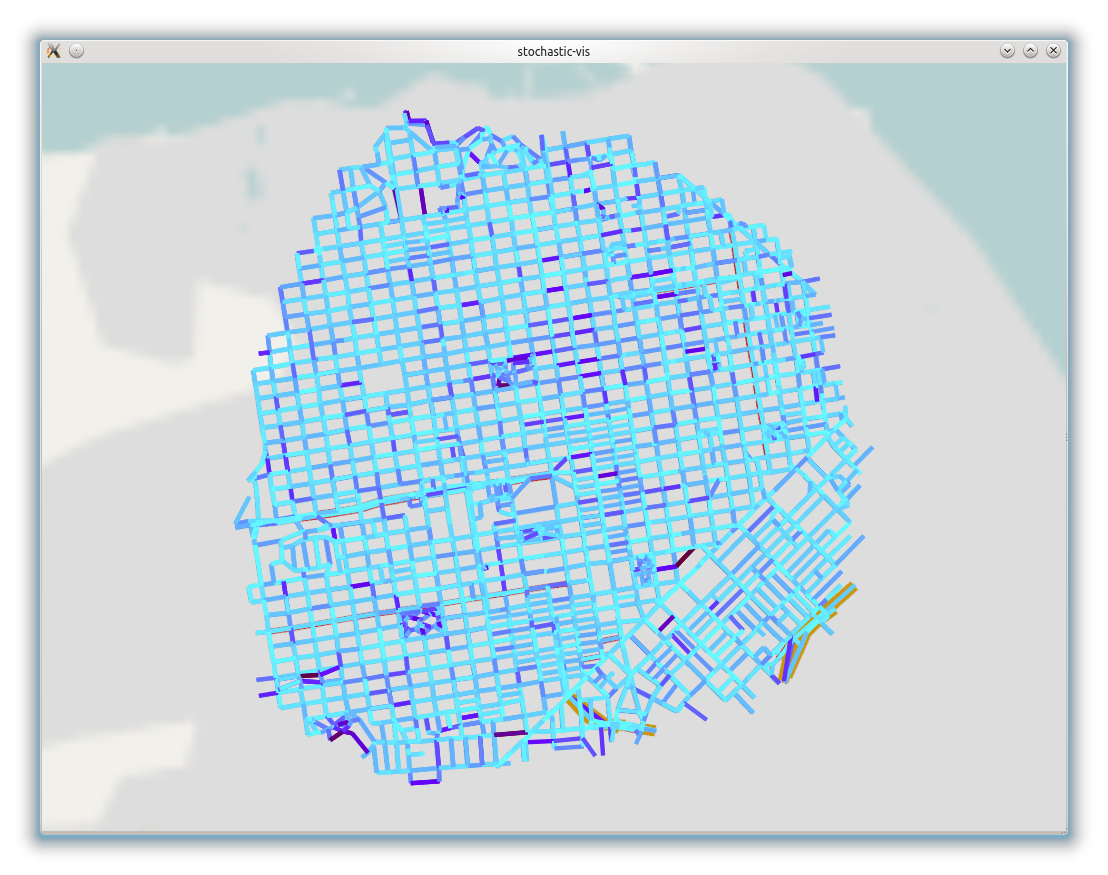}}\hfill
	\subfigure[Network]{\includegraphics[width=0.47\textwidth]{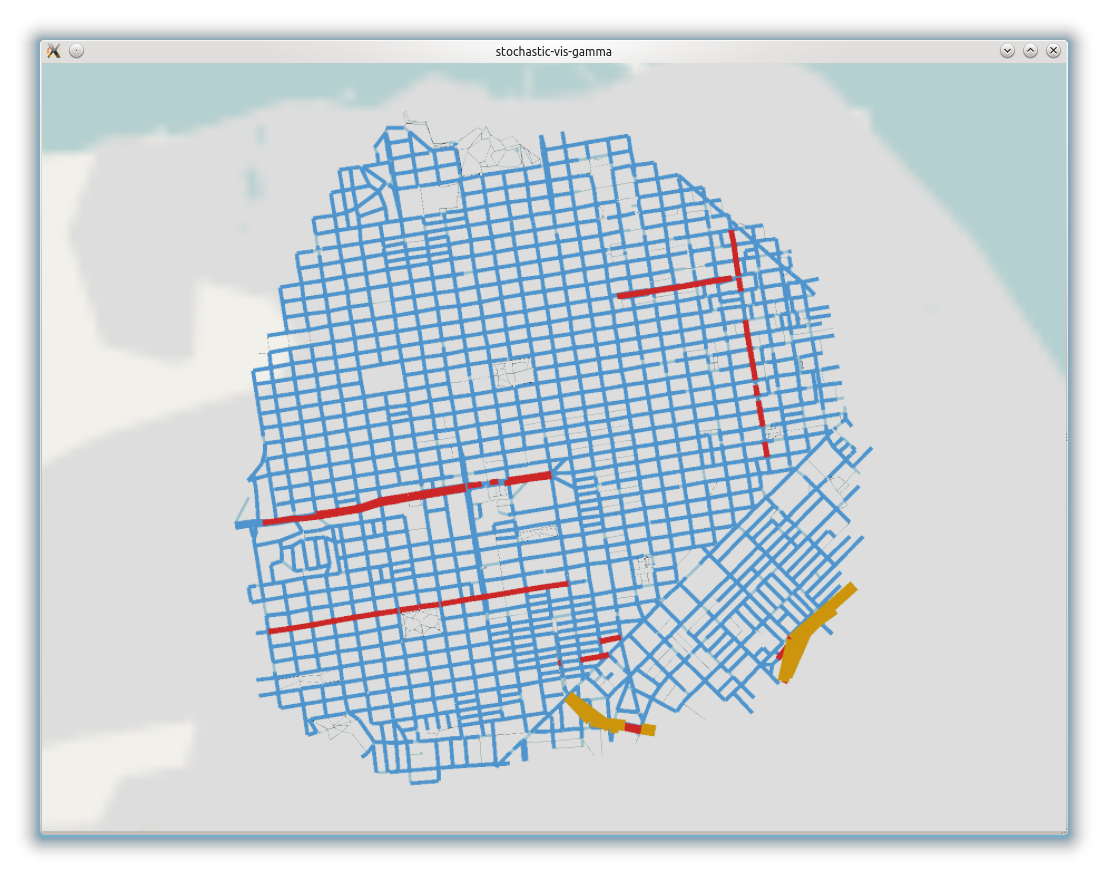}}
	\caption{Illustration of Arcs with high variance on the network. The darker the color of an arc, the more variance is associated with it. Given in relative scale from 0~\% (light blue) to 100~\% (dark red). Graph names are from Table~\ref{tab:input-char}.}
	\label{fig:variance-network-2}
\end{figure}

\end{appendix}
\end{document}